\input{epsf}
\documentclass[prb,aps,twocolumn,showpacs]{revtex4}
\usepackage{amssymb}
\usepackage{amsfonts}
\usepackage{epsfig}

\begin{document}
\title{Percolation model for structural phase transitions in 
Li$_{1-x}$H$_x$IO$_3$ mixed crystals }

\author{P. H. L. Martins$^{1,2}$\footnote{Email: pmartins@uft.edu.br}, 
J. A. Plascak$^{2}$\footnote{Email: pla@fisica.ufmg.br}, 
and M. A. Pimenta$^{2}$\footnote{Email: mpimenta@fisica.ufmg.br} }

\affiliation{$^1$Universidade Federal do Tocantins, 
Caixa Postal 111, 77001-970, Palmas, TO - Brazil \\ 
$^2$Departamento de F\'{\i}sica, Instituto de Ci\^encias Exatas, 
Universidade Federal de Minas Gerais, Caixa Postal 702, 30123-970, 
Belo Horizonte, MG - Brazil }
\date{\today}

\begin{abstract}
A percolation model is proposed to explain the structural phase 
transitions found in Li$_{1-x}$H$_x$IO$_3$ mixed crystals 
as a function of the concentration parameter $x$. 
The percolation thresholds are obtained from Monte Carlo
simulations on the specific lattices occupied by lithium atoms and 
hydrogen bonds. The theoretical results strongly suggest that 
percolating lithium vacancies and hydrogen 
bonds are indeed responsible for the solid solution observed in 
the experimental range $0.22 < x < 0.36$.

\end{abstract}

\maketitle

The lithium iodate ($\alpha$-LiIO$_3$, hexagonal lattice) and the 
iodic acid ($\alpha$-HIO$_3$, orthorrombic lattice) have been both 
intensively investigated, owing  to their interesting properties and 
useful optical applications \cite{Kurtz68,Nash69,Coquet83,Svensson83}.
The lithium iodate is also interesting for fundamental reasons, 
specially those related to its structural phase transitions 
\cite{Crettez87}. It is known that most of the properties of the 
$\alpha$-LiIO$_3$ are strongly affected by its growth conditions, 
related to three mainly factors: the temperature, the evaporation rate, 
and the pH of the mother LiIO$_3$--H$_2$O--HIO$_3$ solution, which 
controls the quantity of impurities in the lattice. In this case, the 
impurities can be hydrogen atoms. Mixed crystals are easier grown using 
more acid solutions (with a greater HIO$_3$ concentration), but they have 
worse optical quality. Li$_{1-x}$H$_x$IO$_3$ mixed crystals are also 
important for technical applications, since they present a high 
piezoelectric coefficient \cite{Hamid77a}.

Although the structures of iodic acid and lithium iodate are known 
for more than sixty years, the structure of the mixed lithium iodate-iodic 
acid solid solution has been refined only in the last few years 
\cite{LeRoy90,Crettez93,LeRoy95}. Regarding the lithium substitution, 
neutron diffraction studies have shown that the hydrogen does not 
substitute the position of the lithium. Instead, when lithium goes out 
of the lattice, its site becomes empty, while the hydrogen forms hydrogen 
bonds between oxygens that belong to different iodate groups \cite{LeRoy95}.

Ricci and Amron \cite{Ricci51} first mentioned the existence of the
so-called solid solution of Li$_{1-x}$H$_x$IO$_3$ for $x$ varying 
continuously over the range 0.22 to 0.36. Since then, much
effort has been dedicated in studying these compounds, but 
the reason of these particular values is not yet well understood.
In this work we propose a percolation model that explains the lower 
and upper limits experimentally observed for crystallizing the solid 
solution of these mixed crystals.

Let us first discuss the principal properties of pure lithium iodate.
Structural phase transitions on this crystal have been largely studied
\cite{Crettez87,Melo82,Peyrard75,Misset76}. At room temperature 
and low pressure, two forms are possible: $\alpha$ (hexagonal) and $\beta$ 
(tetragonal). The $\alpha$ structure has been known since 1931 
\cite{Zachariasen31} and is the more stable one at room temperature. 
There are two molecules per unit cell with the lithium atoms disposed 
on a simple hexagonal lattice. Lattice constants in \AA, with errors in 
parenthesis, are given by \cite{Coquet83}: $a$ = 5.48(1) and 
$c$ = 5.18(1). The Li--O distance is approximately 2.12 \AA.
As the temperature increases, $\alpha$-LiIO$_3$ exhibits 
two first-order phase transitions: $\alpha \rightarrow \gamma 
\rightarrow \beta$, where phases $\gamma$ and $\beta$ have 
orthorhombic and tetragonal symmetries, respectively \cite{Crettez87}. 
Several works reveal discrepancies in the transition temperatures 
as well as in the temperature range where the $\gamma$ phase exists. 
These differences are attributed to, among other reasons, the quantity 
of HIO$_3$ impurities in the lattice 
\cite{Melo82,Peyrard75,Misset76}. Thus, a good understanding 
of the alterations caused by the presence of hydrogen atoms is essential 
for studying these phase transitions.

\begin{figure}[htbp]
\centering
\vspace{0.2in}
\epsfxsize=7cm
\epsfysize=7cm
\epsfbox{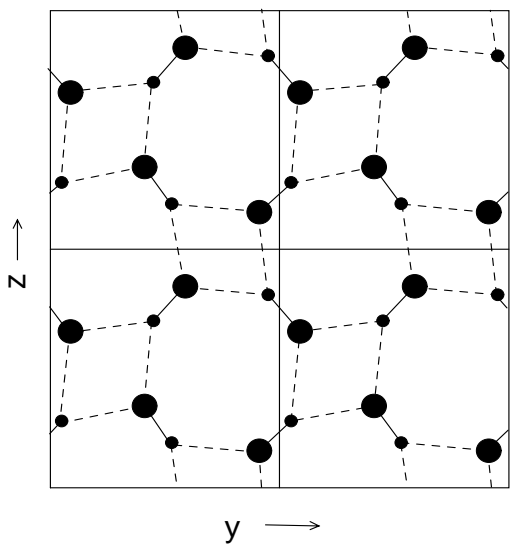}
\vspace{0.2in}
\caption{Schematic representation of the HIO$_3$ structure onto the 
         $yz$ plane. Bigger circles represent IO$_3$ groups and the 
	 smaller ones hydrogen atoms. Bifurcated hydrogen bonds 
	 are represented by solid lines which bifurcate on dashed lines
         at the hydrogen atoms (after Ref. \cite{Rogers41}).}
\label{Fig1}
\end{figure}

Let us now turn to the $\alpha$-HIO$_3$ crystal.
The structure of $\alpha$-iodic acid, though not as simple as that 
of the $\alpha$-LiIO$_3$, has been known since 1941, by means of 
X-ray diffraction\cite{Rogers41}. It is orthorhombic, with 
four molecules per unit cell and lattice constants in \AA: 
$a$ = 5,520(5); $b$ = 5,855(5) and 
$c$ = 7,715(5). The iodate group is pyramidal, although slightly 
distorted, and has the following interatomic distances in \AA 
(with error of 0.04 \AA\ or less): I--O$_{\rm I}$ = 1.81; 
I--O$_{\rm II}$ = 1.89; I--O$_{\rm III}$ = 1.80; 
O$_{\rm I}$--O$_{\rm II}$ = 2.75; O$_{\rm I}$--O$_{\rm III}$ = 2.78 
and O$_{\rm II}$--O$_{\rm III}$ = 2.78, where O$_{\rm I}$, O$_{\rm II}$ 
and O$_{\rm III}$ are the three oxygens of the iodate group.
The hydrogen is bonded to the oxygen O$_{\rm II}$, at a distance of 1.01 \AA. 
This oxygen has two other neighbors: O$^{\prime}_{\rm I}$ and 
O$^{\prime \prime}_{\rm III}$. Here, prime and double primes denote 
different iodate groups. 
O$_{\rm II}$ forms two hydrogen bonds, O$_{\rm II}$--H--O$^{\prime}_{\rm I}$ 
and O$_{\rm II}$--H--O$^{\prime \prime}_{\rm III}$, of equal intensities. 
The distances between the two oxygens indicate that the hydrogen bond
is strong (there are here two bonds per iodate group each of which is about 
the strenght of the single hydrogen bond in water and ice) \cite{Rogers41}.
A better way to visualize this structure is to look at
a projection onto the $yz$ plane, as shown schematicaly in Fig. \ref{Fig1},
where the whole IO$_3$ group has been represented by just one symbol
(and no distinction is made among the three oxygens).
There exists a kind of {\it bifurcated} hydrogen bond: a single proton 
is able to form two hydrogen bonds, i.e., the hydrogen closest to an 
iodate forms a bifurcated bond with two other iodates. Thus, there are 
two hydrogen bonds for each iodate group. These bifurcated bonds and the 
weak bonds between iodine and oxygens belonging to other iodate groups 
hold the iodates together.

The existence of bifurcated hydrogen bonds has been known since 
1939, when the structure of glycine was determined 
\cite{Albrecht39}. Nevertheless, with the exception of a few 
early studies \cite{Newton79,Newton83,Giguere87}, this subject 
has not received much attention. The term {\it bifurcated} is used 
to denote two different kinds of bond. 
In a bifurcated {\it acceptor}, two different acceptor oxygens share 
the same donated proton. In the case of a bifurcated {\it donor} 
a single oxygen donates two protons to another single oxygen. 
The first case is what occurs in the structure of iodic acid, 
as determined in Ref. \cite{Rogers41}.

\begin{figure}[htbp]
\vspace{0.2in}
\epsfxsize=8cm
\epsfysize=6cm
\epsfbox{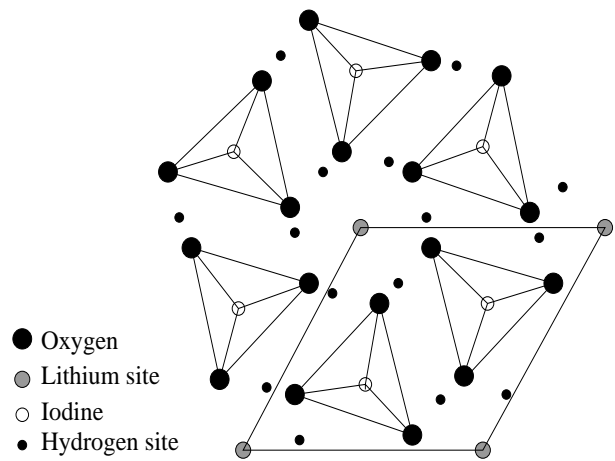}
\vspace{0.2in}
\caption{The crystal structure of Li$_{1-x}$H$_x$IO$_3$ ($x \approx 0.33$) 
projected onto the $xy$ plane, according Ref. \cite{LeRoy90}. In this case, 
bifurcated bonds are not considered. The protons do not 
replace the lithium ions, but are randomly distributed in general positions 
between two oxygen atoms belonging to different iodate groups.}
\label{Fig2}
\end{figure}

Several works have been devoted to the mixed crystal Li$_{1-x}$H$_x$IO$_3$ 
\cite{LeRoy90,Ricci51,Hamid77b,Avdienko85,Barabash90,Pimenta97}.
Ricci and Amron \cite{Ricci51} first reported the existence of the
solid solution for hydrogen concentrations $x$ in the range $0.22 < x < 0.36$.
In 1990, Le Roy {\it et al}. \cite{LeRoy90} concluded that, for $x=0.33$,
the mixed crystal possesses an average hexagonal symmetry, with an 
arrangement close to the HCP (hexagonal close-packed). The structure 
at room temperature is shown in Fig. \ref{Fig2}. Indeed, they 
observed that protons do not substitute the lithium atoms (2a in 
Wyckoff notation), but are randomly distributed through the 
lattice, occupying general positions (6c) 
between two oxygen atoms that belong to different iodates.
Raman investigations \cite{Pimenta97} are consistent with this structure 
and show that the random presence of hydrogens changes the selection rules 
for the Raman scattering for $x$ in the range $0.27 < x < 0.36$.
The lattice constants are found to vary linearly with the 
concentration $x$, according the relations \cite{LeRoy90}: 
$a =  5.464 + 0.282x$ and $c = 5.165 - 0.622x$, with $a$ and $c$ in \AA.  
So, it is possible to obtain $x$ via: $x = 5.985 - 6.335 c/a$, 
valid for $0.22 < x < 0.36$.

As mentioned before, only in the range $0.22 < x < 0.36$ the solid 
solution is constituted as a single phase. Out of this range, 
there is a mixture of solid solution and the pure compounds 
(LiIO$_3$ and HIO$_3$). Thus, these limits seem to be favorable for 
crystallizing the mixed crystal. Our purpose is identifying these values 
as the percolation thresholds on the corresponding lattices. 
In this manner, we expect that 0.22 correspond to the critical point
for site percolation on the LiIO$_3$ (simple hexagonal) lattice. 
Similarly, 0.36 might correspond to the percolation threshold for bond 
percolation on the HIO$_3$ (orthorhombic) lattice. This seems to be 
plausible since, as has been discussed above, it has been observed that 
the protons do not substitute the lithium atoms, but are randomly 
distributed through the lattice, occupying general positions between two 
oxygen atoms that belong to different iodates \cite{LeRoy90,Pimenta97}. 
The proton positions should obey the arrangement depicted in Fig. 1. As 
the proton mediates the hydrogen bond, the net effect is a random 
dilution of hydrogen bonds on the crystal structure of Fig. 1. Thus, 
this system turns out to be a quite interesting problem involving site 
and bond dilution in the same crystalline structure.

It is known that, when lithium atoms are removed from their 
sites on the original 
hexagonal lattice, vacancies are left on their positions \cite{LeRoy90}. 
As more lithium atoms go out (which corresponds to an increasing $x$), 
clusters of neighboring vacancies are formed. There exists a given 
critical concentration $x_c$ at which the lithium vacancies percolate 
through the lattice. In other words, for vacancies concentrations $x$ 
such that $x \ge x_c$, there are an infinite cluster of neighboring 
vacancies. This fact should means that, for $x \ge x_c$, the presence 
of vacancies causes an instability on the LiIO$_3$ hexagonal lattice
and then propitiates the crystallization of the solid solution.

On the other hand, concomitantly to the removal of lithiums, hydrogen 
atoms are added in the lattice, originating H-bonds. By an analogous 
argument, when the number of H-bonds is sufficiently large, there will 
be a cluster of neighboring bonds that percolates through the lattice, 
being possible to consolidate and stabilize the HIO$_3$ orthorhombic lattice.

We performed Monte Carlo simulations in order to obtain the percolation
thresholds on the corresponding lattices. The algorithm we have used is 
that due to Newman and Ziff \cite{Newman00}, applied to three-dimensional 
lattices \cite{Puli}. Despite to the fact that the percolation threshold 
for the hexagonal lattice need not be recomputed, since its value is quite 
well known from the literature [for instance, $p_c = 0.2625(2)$ \cite{Puli}, 
$p_c = 0.2623(2)$ \cite{Marck97}], we have done just one simple simulation 
on a large lattice to see the performance of the method in order to apply it 
to the less studied case of the present bond dilution. In the site dilution 
problem, for a given system size, 
we start with an empty lattice. So, we fill the lattice, choosing sites 
to be occupied, at random. When a new site is added, it can originate an 
isolated cluster (if all its neighbors are empty), or it joins together
two or more clusters. After a complete filling of the lattice (which 
corresponds to one Monte Carlo step), we can evaluate quantities 
of interest, like the percolation threshold. Repeating this procedure 
several times, we get more accurate results. Using periodic boundary 
conditions, the percolation threshold is the concentration of occupied 
sites at which, for the first time, a cluster of neighboring occupied 
sites wraps around the system. The bond problem is treated in an
analogous way.

For simulating the removal of the Li atoms, we can think that it is 
necessary to begin with a full lattice and then take out sites at random. 
In this sense, percolation means that there exists an infinite cluster of 
vacancies. If we replace {\it Li atoms} by {\it empty sites} and 
{\it vacancies} by {\it occupied sites}, we restore the standard
percolation problem.

We applied the cited algorithm on the simple hexagonal lattices with up to 
8~000 sites, which corresponds to a linear dimension $L=20$. 
As we are interested only in the lithium positions, each site represents a 
lithium atom (the iodates localization are not relevant in this case). 
Performing $2.0 \times 10^5$ Monte Carlo steps, one obtains $x_c = 0.26$ 
for the percolation threshold on this lattice. It is clear from the above 
result that for the purpose of the present study just this lattice gives a 
value which is quite comparable to those listed above from a more accurate 
finite-size scaling approach, namely, $p_c = 0.2625(2)$ \cite{Puli} 
and $p_c = 0.2623(2)$ \cite{Marck97}. Within our conjecture, 
this value might correspond to the lower limit for crystallizing the 
solid solution of Li$_{1-x}$H$_x$IO$_3$.

Due to the complex structure of HIO$_3$, some modifications on the 
algorithm concerning the bond dilution are needed, in comparison
with the usual percolation model. Since hydrogen bonds are formed 
between different iodate groups, we can represent, in our simulations, 
each iodate by a single site in the lattice. One has the restriction 
that bonds can exist only between sites that are located on different planes 
in order to reflect the fact that, in the real iodic acid, the distance 
between neighboring iodates on the same plane is greater than the maximum 
value for existing hydrogen bonds. The presence of bifurcated bonds leads to 
the additional constraint that one has always to choose a couple of bonds 
together.
Since we know {\it a priori} the bonds that share the same hydrogen, 
we can choose a site (iodate) and then fill the two corresponding bonds. 
Using lattices with a total number of sites up to 6~912 (which corresponds 
to a linear size $L=12$, since there are four molecules per unit cell), 
and $5.0 \times 10^5$ Monte Carlo steps for each system size, 
simulations yield $x_c = 0.33$ for the critical hydrogen concentration 
on the specified crystal structure. Assuming the same accuracy as the 
previous site diluted model we believe that the number above is quite fine 
to be compared with the experimental one. A finite-size scaling estimate, 
providing a critical concentration with more digits of precision, would 
be irrelevant for the present purpose.

Before achieving these results, other similar hypotheses were tested, 
although all of them sharing the same fundamental concepts. 
The first attempts did not include bifurcated bonds. Bonds were choosen 
randomly one by one. Besides results were unsatisfactory, there is an 
incompatibility between the concentration of lithium vacancies and the 
corresponding concentration of hydrogen bonds. This can be
explained as follows. For $x = 0$ (pure LiIO$_3$) there is one lithium for 
each iodate. Imagine a lattice with a total of $N$ lithium atoms. 
So, there are also $N$ iodates. Since each iodate has four neighbors 
there are $2N$ possible bonds. Removing one lithium, we need to introduce 
one hydrogen. If each hydrogen forms a single bond, when all $N$ lithium 
atoms will be removed ($x = 1$), there will $N$ hydrogen bonds, which 
corresponds to a fraction $N/2N = 0.5$ of the total number of bonds. 
This fact precludes the equivalence between the hydrogen concentration 
$x$ and the related percolation thresholds. Bifurcated bonds avoid this 
problem, since $N$ hydrogens form $2N$ bonds. Thus, the fraction of 
vacancies is equal to the fraction of bonds.

According to our results, the solid solution should exist as a 
single phase for $x$ in the range $0.26 < x < 0.33$. In this sense, 
when $x$ rises from 0 to 0.26, lithium atoms are removed from the 
lattice but the $\alpha$-LiIO$_3$ structure remains stable. 
For $x > 0.26$, the number of vacancies is so great 
that the $\alpha$-LiIO$_3$ lattice does not support itself. 
Concomitantly, while $x < 0.33$, the hydrogen bonds are insufficient 
for supporting the HIO$_3$ orthorhombic structure. Thus, in the range 
$0.26 < x < 0.33$ the system is disordered and we find the so-called 
solid solution. For $x > 0.33$ the percolating cluster of bonds 
maintains the HIO$_3$ lattice stable.

We conclude that our results are in good agreement with the experimental 
values and, although not stated in the experimental works, within the 
experimental errors. [For instance, the smaller concentration varies 
from 0.22 \cite{Ricci51} to 0.27 \cite{Pimenta97}]. Thus, our results 
evidence the percolating character on these mixed compounds and 
clearly indicate that the crystal stability is, in fact, related to the 
percolation thresholds supported by the corresponding crystal structures.
Further experimental studies regarding the structural transition of this 
material would be very welcome, mainly concerning its second-order character.

Fruitful discussions with R. Dickman and G.J.M. Garcia, and D.P. Landau are 
gratefully acknowledged. This research was supported in part by the Brazilian 
agencies CNPq, CAPES and FAPEMIG.

\end{document}